\documentclass{aa}
\usepackage{graphicx}
\begin{document}


%
\def\EE#1{\times 10^{#1}}
\def\gcm{\rm ~g~cm^{-3}}
\def\cm3{\rm ~cm^{-3}}
\def\kms{\rm ~km~s^{-1}}
\def\cms{\rm ~cm~s^{-1}}
\def\ergs{\rm ~erg~s^{-1}}
\def\wl{~\lambda}
\def\wll{~\lambda\lambda}
\def\Nii{M(^{56}{\rm Ni})}
\def\FeI{{\rm Fe\,I}}
\def\FeII{{\rm Fe\,II}}
\def\FeIII{{\rm Fe\,III}}
\def\Niii{M(^{57}{\rm Ni})}
\def\FeIb{{\rm [Fe\,I]}}
\def\FeIIb{{\rm [Fe\,II]}}
\def\FeIIIb{{\rm [Fe\,III]}}
\def\Msun{~{\rm M}_\odot}
\def\Ti44{M(^{44}{\rm Ti})}
\def\MZA{M_{\rm ZAMS}}
\def\mum{\mu{\rm m}}

\def\lsim{\!\!\!\phantom{\le}\smash{\buildrel{}\over
  {\lower2.5dd\hbox{$\buildrel{\lower2dd\hbox{$\displaystyle<$}}\over
                               \sim$}}}\,\,}
\def\gsim{\!\!\!\phantom{\ge}\smash{\buildrel{}\over
  {\lower2.5dd\hbox{$\buildrel{\lower2dd\hbox{$\displaystyle>$}}\over
                               \sim$}}}\,\,}

\title{ISO/SWS observations of SN 1987A: II. A refined upper limit on the 
mass of $^{44}$Ti in the ejecta of SN 1987A.\thanks{ISO is an ESA project with
instruments funded by ESA Member States (especially the PI countries: France,
Germany, the Netherlands and the United Kingdom) and with the participation of 
ISAS and NASA.}}

\author{Peter Lundqvist\inst{1} 
\and Cecilia Kozma\inst{1} 
\and Jesper Sollerman\inst{1,2} 
\and Claes Fransson\inst{1}
}
\institute{Stockholm Observatory, SE-133 36 Saltsj\"obaden, Sweden.
\and European Southern Observatory, Karl-Schwarzschild-Strasse 2, 
D-85748 Garching bei M\"unchen, Germany
}

\date{Received:  Accepted: }

\mail{peter@astro.su.se}

\titlerunning{Mass of $^{44}$Ti in the ejecta of SN 1987A}
\authorrunning{P. Lundqvist et al.}

\abstract{
ISO/SWS observations of SN 1987A on day 3\,425 show no emission in 
[Fe~I]~24.05~$\mum$ and [Fe~II]~25.99~$\mum$ down to the limits $\sim 
0.39$~Jy and $\sim 0.64$~Jy, respectively. Assuming a homogeneous 
distribution of $^{44}$Ti inside $2\,000 \kms$ and negligible dust cooling, 
we have made time dependent theoretical 
models to estimate an upper limit on the mass of ejected $^{44}$Ti. 
Assessing various  uncertainties of the model, and checking the late optical 
emission it predicts, we obtain an upper limit of $\simeq 1.1\EE{-4}\Msun$. 
This is lower than in our previous estimate using other ISO data, 
and we compare our new 
result with other models for the late emission, as well as with expected yields 
from explosion models. We also show that steady-state models for the optical 
emission are likely to overestimate the mass of ejected $^{44}$Ti.
The low limit we find for the mass of ejected $^{44}$Ti could be higher if dust 
cooling is important. A direct check on this is provided by the 
gamma-ray emission at 1.157 Mev as a result of the radioactive decay 
of $^{44}$Ti.
}

\maketitle

\section{Introduction}
The late ($t \gsim 150$~days) emission from supernova (SN) 1987A probes 
the content of radioactive isotopes in the supernova ejecta. Due to the proximity 
of the supernova, it has been possible to follow its evolution
to very late stages when $^{44}$Ti takes over from $^{56}$Co and $^{57}$Co 
as the most important radioactive isotope for the powering of the light curve, 
i.e., after $\sim 1\,500 - 2\,000$ days (Woosley et al. 1989; 
Kumagai et al. 1991; Kozma \& Fransson 1998a; Kozma 2000, henceforth K00). 
Both $^{56}$Co and $^{57}$Co are decay products of more short-lived nickel 
isotopes with the same mass numbers. 
Measuring the radioactive input therefore provides 
estimates of the ejected masses of $^{56}$Ni, $^{57}$Ni and $^{44}$Ti, 
i.e., $\Nii$, $\Niii$ and $\Ti44$, respectively, which constrain models of 
the explosion (e.g., Woosley \& Weaver 1995; Thielemann et al. 1996; 
Nagataki et al. 1997).

While $\Nii$ and $\Niii$ are rather accurately known for SN 1987A, 
with $\Nii \approx 0.071 \Msun$ (e.g., Suntzeff \& Bouchet 1990),
and $\Niii \sim 3.3\EE{-3} \Msun$ (Fransson \& Kozma 1993, and references
therein), the situation is more uncertain for $^{44}$Ti. To some extent this 
has been due to a poorly known decay time of this isotope, although this has
recently improved; $^{44}$Ti decays first to $^{44}$Sc, on a time scale 
of $87.0\pm1.9$ years (Ahmad et al. 1998; G\"orres et al. 1998; see also
Norman et al. 1998 who obtain $86\pm3$ years), and then quickly further 
to $^{44}$Ca.

Based on time dependent modeling, K00 estimates
that $\Ti44 = (1.5 \pm 1.0) \times 10^{-4} \Msun$ best explains the evolution
of the optical broad-band photometry of SN 1987A for $t \lsim 3\,270$ days. 
This is consistent with Chugai et al. (1997), who 
find $\Ti44 \sim (1-2)\EE{-4} \Msun$ from the optical line emission 
at 2\,875 days, and Mochizuki \& Kumagai (1998) 
who obtain the same result from a light curve analysis (see also Nagataki 
2000). However, as is shown by K00, both broad-band photometry and the 
emission modeled by Chugai et al. (1997) constitute only a minor fraction 
of the total emission put out by the supernova. At these epochs most of the 
emission from the supernova instead comes out in the far infrared (IR)
in a few iron lines, most notably [Fe~II]~25.99~$\mum$. 
This makes bolometric corrections to the late 
optical data rather uncertain, and a more direct way to 
measure the content of $\Ti44$ is to measure the flux in the far-IR 
lines. 

In a recent study Lundqvist et al. (1999; henceforth L99) analyze 
Infrared Space Observatory (ISO; Kessler et al. 1996) data obtained mainly 
at $t = 3\,999$ days. L99 couple these observations to time dependent
model calculations similar to those in K00, 
and from the absence of iron line emission in the
ISO spectra they derive $\Ti44 \lsim 1.5\EE{-4} \Msun$. This limit is 
consistent with the results of Chugai et al. (1997), Mochizuki \& Kumagai 
(1998) and K00. A different limit was obtained by Borkowski et al. (1997) 
who in their preliminary analysis of ISO data from $t = 3\,425$ days 
derive $\Ti44 \lsim 1.5\EE{-5} \Msun$. Such a low titanium mass would have 
important consequences for models of the explosion. Here we analyze the data 
of Borkowski et al. in the same way as was done for the $ 3\,999$ day data in 
L99, however, expanding the discussion on the uncertainties of our 
method. In particular, we discuss the uncertainty of the temperature, 
which is important since the far-IR lines are collisionally excited, and 
their excitation energy (e.g., $T \approx 550$~K for the 26~$\mum$ line) 
is much higher than the gas temperature. Because the emission at this 
epoch is powered mainly by positrons from the decay of $^{44}$Ti, we also 
discuss the uncertainty of the line fluxes due to the efficiency of the 
trapping of positrons.

\section{Observations and Results}

We have used the ISO data discussed in Borkowski et al. (1997). These data
were obtained on 10 July, 1996 (i.e., on day 3\,425 of SN 1987A)
in the SWS06 mode of the Short Wavelength Spectrograph (SWS; de Graauw et al.
1996). The data we
analyze span the regions $23.59  - 25.04~\mum$ and $25.51 - 26.53~\mum$. 
Both these ranges are covered by band 3D of SWS, and include the lines
[Fe~I]~24.05~$\mum$ and [Fe~II]~25.99~$\mum$, i.e., the two strongest lines 
expected from the supernova (cf. L99). These data are superior to 
those discussed in L99 as the SWS06 scans are deeper than the 
SWS01 scans in L99, and include a large enough wavelength interval for 
accurate continuum determination compared to the SWS02 data in L99.

The data were retrieved from the ISO archive, and have been reduced 
with the off-line processing (OLP/pipeline) version 8.7. We used ISO 
Spectral Analysis Package (ISAP) to flatfield and average over the scan 
directions. The averaging over the detectors was done 
with a resolution of R = 300, using the standard $2.5\sigma$-clipping. 

\begin{figure}
\hskip 0.1cm \resizebox{8.5cm}{!}{\includegraphics{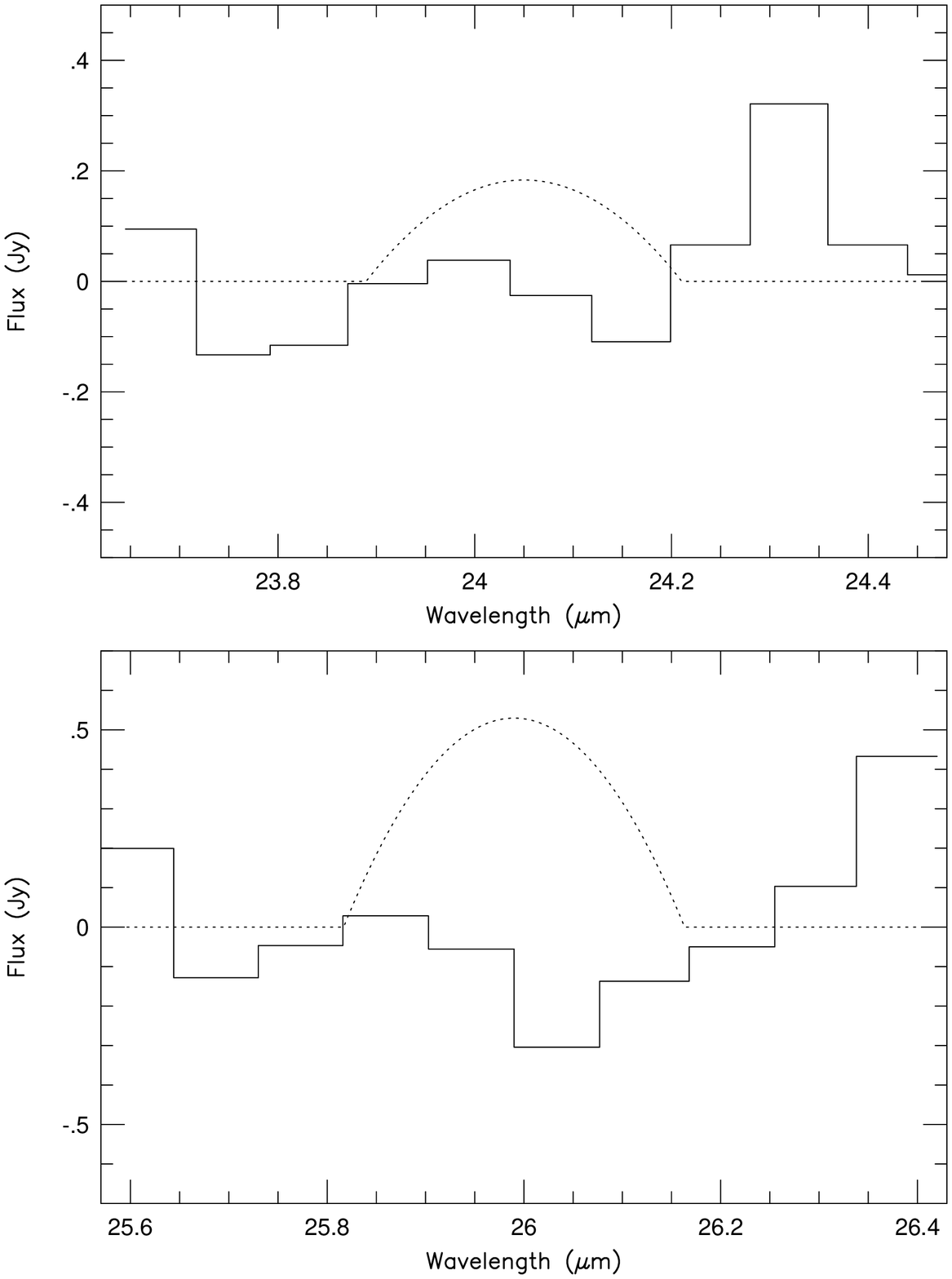}}
\caption{
  ISO SWS/AOT6 spectra of SN 1987A on day 3\,425. The spectra 
  were reduced using the pipeline software (see text). The bin size is 
  $\approx 0.081~(0.086)~\mum$ for [Fe~I]~24.05~$\mum$ ([Fe~II]~25.99~$\mum$)
  (see text), which is low enough compared to the expected line width. 
  Overlaid (dashed lines) are the line emissions from a model 
  with $\Ti44 = 5\EE{-5} \Msun$ (without photoionization) described
  in Table 1 and Sect. 3.2. To obtain the modeled line profiles we have 
  assumed a maximum core velocity of $2\,000 \kms$, and a homogeneously
  distributed emission throughout the core (see Sect. 3.2). The peak flux
  of [Fe~II]~25.99~$\mum$ for this model is slightly less than what is 
  needed for a 3$\sigma$ detection (cf. Sect. 3.3).
  }
\label{observed.ps}
\end{figure}

In Fig. 1 we present fully reduced spectra of SN 1987A for the two 
wavelength regions. Although the nominal instrumental resolution of the SWS 
spectra is R~$\approx 1\,800$ we have averaged the spectra with a bin size 
of R = 300, corresponding to $\approx 1\,000 \kms$. This is sufficient 
to resolve the lines since they could extend to well over $2\,000 \kms$. 
However, we detected neither [Fe~I]~24.05~$\mum$ nor [Fe~II]~25.99~$\mum$ in 
the spectra. We estimated the RMS in the data with a zero-order 
baseline fit in the regions $23.80  - 24.80~\mum$ and $25.60  - 26.40~
\mum$, and found $0.11$~Jy and $0.19$~Jy, respectively. 

The exact RMS-values vary somewhat with the bin size used in the 
averaging routine. Reducing the bin size to R = 500 gives 
the limits $0.13$~Jy and $0.21$~Jy, respectively. In order to 
derive a conservative limit on the $^{44}$Ti mass we have adopted these 
values. A 3$\sigma$ limit for the $26~\mum$ line then becomes $\sim 
0.64$~Jy, while for the $24~\mum$ line it is lower, $\sim 0.39$~Jy. The bin 
size is small enough so that we can use the limits for the peak of the 
expected line profiles in our estimate of the $^{44}$Ti mass.

\section{Interpreting the SWS observations}

\subsection{The model for line emission from the supernova}
To interpret the upper limits on the iron line fluxes in Sect. 2, we 
have made model calculations similar to those in L99.
Our computer code is described in detail in Kozma \& Fransson (1998a), 
and summarized in L99. Here we only give a very brief 
recapitulation of the model, and discuss recent improvements.

The thermal and ionization balances are solved time-dependently, 
as are also the level populations of the most important ions.
A total of $\sim 6\,400$ lines are included in the calculations.
 
The radioactive isoptopes included are $^{56}$Ni, $^{57}$Ni, and $^{44}$Ti,
and we calculate the energy deposition of gamma-rays and positrons solving the 
Spencer-Fano equation (see Kozma \& Fransson 1992). We assume that the 
positrons deposit their energy locally.

We include $0.07 \Msun$ of $^{56}$Ni and $3.33\EE{-3} \Msun $ 
of $^{57}$Ni (Sect. 1). In L99 three values of $\Ti44$ were 
tested: $\Ti44 =0.5\EE{-4}, 1.0\EE{-4}$ and $2.0\EE{-4} \Msun$. Here we extend
this to include three new values: $\Ti44 = 1.0\EE{-5}, 1.7\EE{-5}$ 
and $3.0\EE{-5} \Msun$. Although we follow the complete evolution of the
supernova after 150 days, we concentrate our discussion mainly on $3\,425$
and $3\,999$ days, i.e., the epochs of the ISO observations.

In L99 an e-folding time of 78 years was used for $^{44}$Ti, as this was 
thought to represent a mean value from experiments. More accurate measurements 
of the decay time (Ahmad et al. 1998; G\"orres et al. 1998; Norman et al. 1998) 
became available while the analysis of L99 was completed, and the results of 
L99 were corrected accordingly before publishing. Here, we use the 
more accurate e-folding time ($87.0\pm1.9$ years) from the outset.

The explosion model we use is the same as in L99
and Kozma \& Fransson (1998a). That is, we take the abundances from the 10H 
model (Woosley \& Weaver 1986; Woosley 1988), but distribute the 
elements so that hydrogen is mixed into the core. 
Spherically symmetric geometry is assumed, and the iron-rich core extends out 
to $2\,000 \kms$, outside of which (out to $6\,000 \kms$) a hydrogen envelope
is attached.

We use the Sobolev approximation for the line transfer. This is a good
approximation for well-separated lines in an expanding medium, but is poorer
when lines overlap. This is actually the case in the UV, especially at 
earlier epochs. 
The overlap leads to UV-scattering which affects the UV-field within the 
ejecta. As in L99, we have studied two extreme cases to test this effect:
photoionization as in the original model, and simply switching off the 
photoionization caused by the UV-field. From the results of L99 we, 
however, do not expect the UV-field to be the dominant source of uncertainty 
in our models. Another drawback of the Sobolev approximation is that we
do not account for line fluorescence in which UV lines are split into optical
and IR lines. To treat this accurately is beyond the scope of this paper, 
but is treated in detail in a forthcoming paper.

Since L99 we have updated the collision strengths for the [Fe~II] $26~\mum$ 
transition. The collision strength, $\Omega$, is given by Zhang \& Pradhan 
(1995) as a function of temperature. The value of $\Omega$ at $T \leq 200$ K 
is set to $\Omega = 7.0$ (A. Pradhan, private communication).
We note, however, that there can still be uncertainty in this result 
as a fully relativistic calculation is needed to accurately calculate 
the collision strength at these low temperatures (M. Bautista, private 
communication).

\subsection{Model calculations}
The results of our calculations can be seen in Table 1. We tabulate the 
fluxes at $3\,425$ days for [Fe~I] $24~\mum$ and [Fe~II] $26~\mum$ 
for the three values of $\Ti44$ discussed in L99: $5\EE{-5}$, $1\EE{-4}$ 
and $2\EE{-4}$ $\Msun$. We also include the three new 
values: $1\EE{-5}$, $1.7\EE{-5}$ and $3\EE{-5}$ $\Msun$. For each value
of $\Ti44$ we tabulate results for models with and without photoionization. 
In all models a simple form of dust absorption was assumed (Kozma \& Fransson
1998b).

\begin{table}
\caption[]{Modeled line flux at 3\,425~days$^{a}$}
\label{modelfluxes}
\[
\begin{tabular}{lccc}
\hline
\noalign{\smallskip}

$\Ti44$ ($\Msun$) & Photoion.$^{b}$ & $f_{24\mum}$ (Jy) & $f_{26\mum}$ (Jy) \\

\noalign{\smallskip}
\hline
\noalign{\smallskip}

$1.0\EE{-5}$ &  yes  &    0.047    &    0.095    \\
$1.0\EE{-5}$ &  no   &    0.051    &    0.074    \\
$1.7\EE{-5}$ &  yes  &    0.072    &    0.17     \\
$1.7\EE{-5}$ &  no   &    0.079    &    0.14     \\
$3.0\EE{-5}$ &  yes  &    0.11     &    0.34     \\
$3.0\EE{-5}$ &  no   &    0.12     &    0.29     \\
$5.0\EE{-5}$ &  yes  &    0.17     &    0.61     \\
$5.0\EE{-5}$ &  no   &    0.18     &    0.53     \\
$1.0\EE{-4}$ &  yes  &    0.26     &    1.37     \\
$1.0\EE{-4}$ &  no   &    0.31     &    1.18     \\
$2.0\EE{-4}$ &  yes  &    0.40     &    2.96     \\
$2.0\EE{-4}$ &  no   &    0.54     &    2.53     \\

\noalign{\smallskip}
\hline
\end{tabular}
\]

$^{a}$Distance = 50 kpc. Homogeneous distribution of emitting gas 
inside $\pm 2\,000 \kms$ has been assumed. The resulting shape of the line 
profile is seen Fig. 1. The flux values are the peak values of the lines 
assuming this profile.\\
$^{b}$Indicates whether or not photoionization has been included in the 
calculation. (See Sect. 3.1.)\\

\end{table}

The line fluxes in Table 1 are for a distance to the supernova of 50 kpc, and
assuming a homogeneous distribution of emitting gas inside $\pm 2\,000 \kms$.
The FWHM velocity for the resulting profile (shown in Fig.~1)
is $v_{\rm FWHM} = 2\,828 \kms$, which is close to what Haas et al. (1990)
observed for [Fe~II] 17.94~$\mum$ at $\sim 400$ days after the
explosion, $v_{\rm FWHM} = (2\,900 \pm 80) \kms$.
 
\begin{figure*}
\centering
\includegraphics{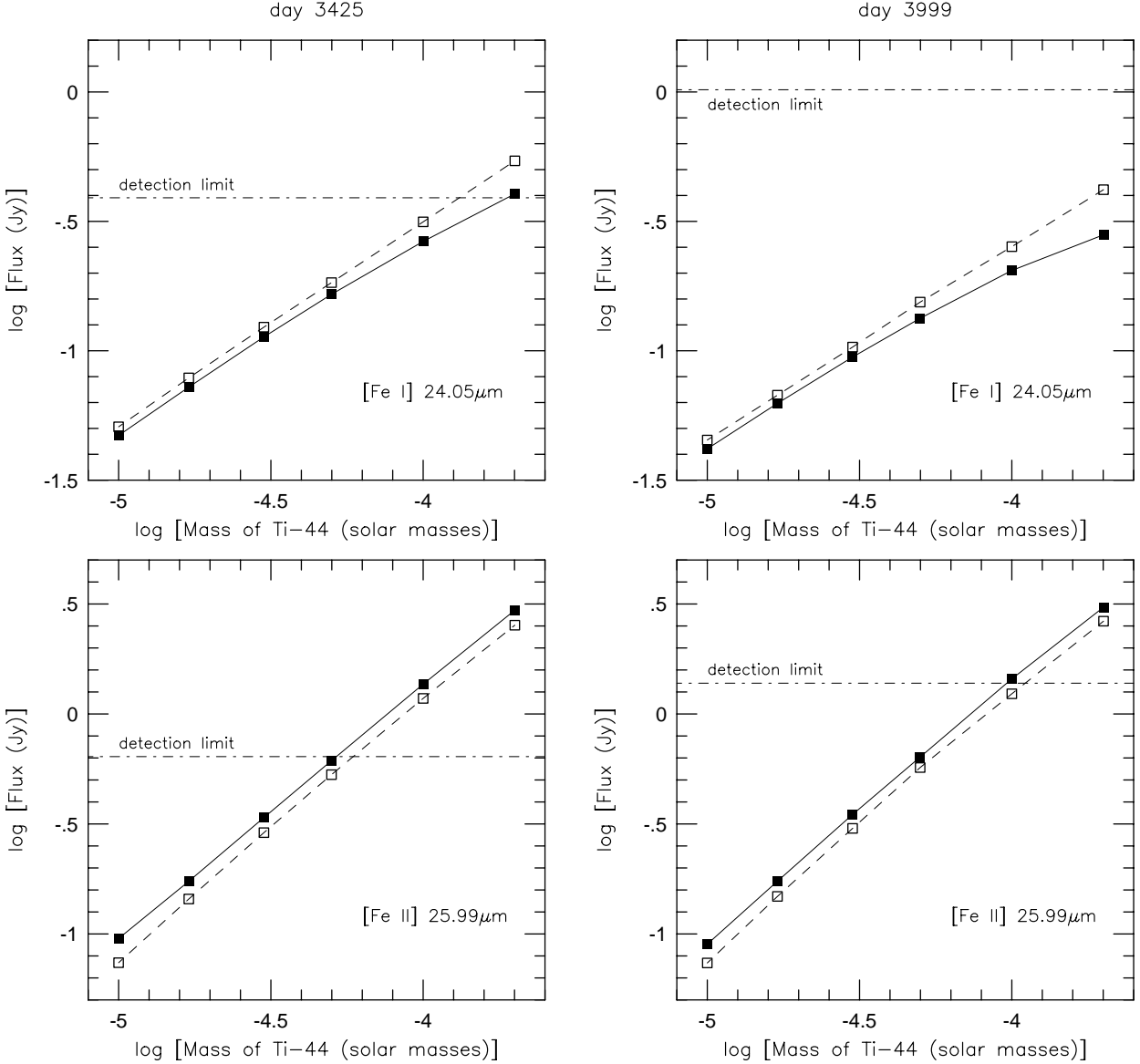}
\caption{Modeled fluxes of [Fe~I] 24.05~$\mum$ and [Fe~II] 25.99~$\mum$ from 
SN 1987A as functions of the mass of ejected $^{44}$Ti, 3\,425 and 3\,999
days after explosion. Models with photoionization included (see text) are 
shown by filled squares and joined by solid lines, while for the others 
photoionization was not included. The inferred $3\sigma$ detection limits 
for the two epochs are shown as dashed-dotted lines. Note the near 
constancy of the modeled fluxes between the two epochs, and the rather small 
uncertainty due to photoionization.
  }
\label{limits.ps}
\end{figure*}

\begin{figure*}
\centering
\includegraphics{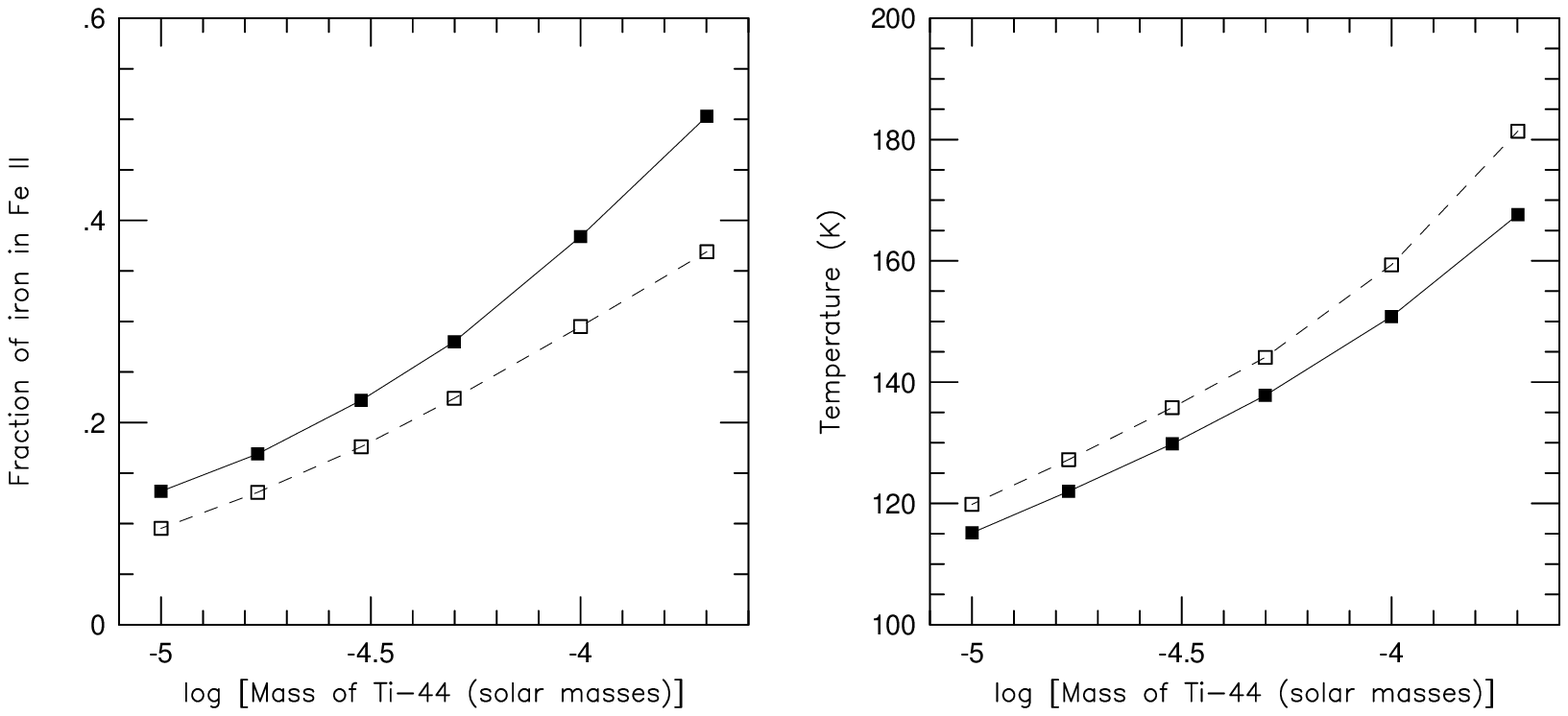}
\caption{Fraction of iron in Fe~II, $X_{\rm FeII}$, and temperature in the
iron-rich part of the ejecta of SN 1987A at 3\,425 days as a function of mass 
of $^{44}$Ti. Symbols and lines have the same meaning as in Fig.~2.
As expected, the temperature and $X_{\rm FeII}$ increase monotonically
with increasing mass of $^{44}$Ti.
  }
\label{structure.ps}
\end{figure*}

In L99 it was discussed how the fluxes in [Fe~I] 24.05~$\mum$ and 
[Fe~II] 25.99~$\mum$ ($f_{24\mum}$ and $f_{26\mum}$, respectively) scale
with increasing values for $\Ti44$ at a given epoch. It was found 
that $f_{26\mum}$ scales nearly linearly with $\Ti44$, while $f_{24\mum}$ 
has a weaker dependence. The stronger dependence for $f_{26\mum}$ is
due to an ionization effect because $X_{\rm FeII}$, the relative fraction of 
iron in Fe~II in the iron-rich gas, increases with increasing $\Ti44$.
We find that this trend continues also for $\Ti44 < 5\EE{-5} \Msun$.
This is shown in Fig.~2 and 3, where models
joined by dashed (solid) lines are without (with) photoionization.
Whereas $X_{\rm FeII}$ and the temperature in the iron-rich gas are only 
shown for 3\,425 days (Fig.~3), 
results for the line fluxes (Fig.~2) are shown for both 3\,425 and 3\,999
days. Note how the fluxes remain nearly constant between these two epochs, 
especially in the case of $f_{26\mum}$. 

The only deviation from a linear log($\Ti44$) versus log($f$) behavior in 
Fig.~2 is for $f_{24\mum}$ at the highest values of $\Ti44$ in the 
models with photoionization. This is because here $X_{\rm FeI}$ 
(the fraction of iron in Fe~I in the iron-rich gas) starts to fall 
significantly below unity (cf. Fig.~3): 
for $\Ti44 = 0.3,(0.5),(1.0),(2.0)\EE{-4} \Msun$, $X_{\rm FeI} \approx 0.78, 
(0.70),(0.58),(0.44)$ at 3\,425 days. (The same number for the case without
photoionization is $X_{\rm FeI} \approx 0.82, [0.76],[0.68],[0.58]$.)
The temperature increases monotonically with $\Ti44$ (Fig.~3) and
affects $f_{24\mum}$ and $f_{26\mum}$ in the same way. This is because
the two lines have nearly the same excitation energy, and their effective 
collision strengths vary only weakly (and in our models are assumed to be
constant over the small temperature regime [$\sim 115-180$~K] found in the 
iron-rich gas in our models). 

Table~1 and Fig.~2 show that the 26~$\mum$ line is 
stronger than the 24~$\mum$ line even 
for the models with the lowest values of $\Ti44$ (and thus the highest 
values of $X_{\rm FeI}$). This is because the collision strength for
[Fe~II] 25.99~$\mum$ line is much larger than for [Fe~I] 24.05~$\mum$.

Table~1 and Fig.~2 also show that switching off photoionization 
does not have a dramatic effect on the line fluxes. Again, the largest 
difference for the fluxes is for the 24~$\mum$ line and at the 
highest $\Ti44$ considered. For values $\leq 5.0\EE{-5} \Msun$, switching
off photoionization only affects $f_{24\mum}$ ($f_{26\mum}$) 
by $\lsim 11\%$ ($\lsim 22\%$) at 3\,425 days. 
The small difference is simply due to the minor shift in the degree of
ionization, from Fe~II to Fe~I, when photoionization is switched off.

\subsection{Mass of $^{44}$Ti from [Fe~II]~25.99~$\mum$}
With the results in Fig.~2 it is straightforward to estimate the upper limit 
on $\Ti44$. It is clear that the best estimate within the framework of our 
modeling comes from the 26~$\mum$ line on day $3\,425$.
Using $f_{26 \mum} \lsim 0.64$~Jy from Sect. 2.1, we find the upper 
limit $\Ti44 \lsim  5.2~(5.9) \times 10^{-5} \Msun$ with (without)
photoionization included. A conservative limit (i.e., the case when 
photoionization is unimportant) is 
therefore $\Ti44 \lsim 5.9 \times 10^{-5} \Msun$.
In Fig. 1 we have included the expected line emission for a model 
with $\Ti44 = 5.0 \times 10^{-5} \Msun$, i.e., close to this limiting mass. 
Although our limit is nearly a factor of four higher than that found by 
Borkowski et al. (1997), using the same data, it is still substantially lower 
than found by L99 from day 3\,999. Using our updated code, the day 3\,999
data only set a limit $\Ti44 \lsim 1.1 \times 10^{-4} \Msun$. This is a 
factor $\sim 1.7$ better than can be obtained from the 24~$\mum$ line on 
day 3\,425. We will evaluate our limit from the 26~$\mum$ line on day 3\,425 
in Sect. 4.1.

\section{Discussion}

\subsection{Uncertainties in the modeling}
There are several uncertainties involved in our modeling of the line fluxes. 
The effect of photoionization was already discussed in the previous section,
and was found to be rather small, as was also argued for in L99. An even 
smaller uncertainty in the estimated $\Ti44$ is introduced by the 
decay time of $^{44}$Ti. Given the small error in this (Ahmad et al. 1998;
G\"orres et al. 1998; Norman et al. 1998), we can safely ignore 
this uncertainty. More uncertain is the distance to the supernova. This 
is uncertain to $\sim 10 \%$ (Walker 1998; Lundqvist \& Sonneborn 2001), 
and a conservative limit is given by Gibson (2000), corresponding to an 
uncertainty in the line flux of $\sim 27 \%$. Because the 
modeled $f_{26 \mum}$ scales almost linearly with $\Ti44$ (Sect. 3.2), 
we have used this number also for the uncertainty in the derived $\Ti44$.  

A source of uncertainty, not investigated in L99, is the rather unknown
rates of charge transfer for the elements in the core. In the models 
in Sect. 3, no charge transfer was included between Fe and excited states 
of He, as well as between atoms and ions of Fe. We have tested the importance 
of including this charge transfer, using the rates suggested by Liu et al.
(1998) in a model with $\Ti44 \sim 1.0 \EE{-4} \Msun$ and no photoionization. 
Compared to the model with the same $\Ti44$ in Sect. 3, the model with the 
full charge transfer has a higher value of $X_{\rm FeII}$ by $\sim 20\%$, 
mainly due to charge transfer between Fe~I and Fe~III. However, the 
temperature in the model with full charge transfer is somewhat lower, 
so the difference in $f_{26 \mum}$ between the two models is 
only $\sim 7 \%$. With full charge transfer included, the estimated $\Ti44$
would therefore be {\it lower} than in Sect. 3, but due to the uncertainty 
of the charge transfer rates, and in order to derive a conservative limit 
on $\Ti44$, we assign a generous error of $10 \%$ in $\Ti44$ due to 
this effect.

L99 studied the uncertainty of the collision strength of the 26~$\mum$ 
line. Although we are now using a better fit to the results 
of Zhang \& Pradhan (1995), the collision strength at very low temperatures 
is still uncertain (see Sect. 3.1). In L99 we assumed an uncertainty 
of $15\%$ in $\Ti44$ due to this, and we retain this number also here.

The collisional excitation of the 24~$\mum$ and 26~$\mum$ lines is more 
sensitive to the temperature (through the ${\rm exp}(-h c/\lambda k T)$ 
term) than to the collision strength. Too high a temperature in our 
models could therefore overestimate the flux in these lines, and we 
would consequently underestimate $\Ti44$. We can check this effect by
studying Table 1 where we see that models with $\Ti44 = 5.0 \EE{-5} \Msun$ 
and $\Ti44 = 1.0 \EE{-4} \Msun$ differ in $f_{26 \mum}$ by a factor 
of $\sim 2.2$ at 3\,425 days. Because the temperature in 
the $\Ti44 = 1.0 \EE{-4} \Msun$ model is $\approx 159~K$ (see Fig. 3), 
a lowering of the temperature by only $\sim 20\%$ would 
decrease $f_{26 \mum}$ to the same level as in 
the $\Ti44 =5.0 \EE{-5} \Msun$ model, for constant $X_{\rm FeII}$.
However, the temperatures in our models are not free parameters but are fixed 
by the heating and cooling. The energy not coming out in the 24~$\mum$ 
and 26~$\mum$ lines must instead come out in other lines (like the optical/IR 
recombination lines) which are less sensitive to temperature, or as emission 
from dust. We will make a consistency check of the optical and far-IR emission 
in our models in Sect. 4.2, but we note already here that only a small 
transfer of energy loss from the far-IR lines to the optical/IR recombination 
lines would increase the optical and IR flux considerably, and then the models 
of K00 for the optical would severely overestimate $\Ti44$. 
We therefore believe we are not making a serious error in the derivation of 
the temperature
in our models. 

Dust emits a continuum which is difficult to detect, but we may for simplicity 
assume that the dust emission does not interact with the gas. In that case, 
dust cooling has exactly the same effect for the 24~$\mum$ and 26~$\mum$ 
line fluxes as just lowering $\Ti44$. (See the models in Sect. 3). 
It may therefore well be that dust cooling could tap 
the supernova of its [Fe~II] 25.99~$\mum$ emission, but it is likely that it 
would do so in a way which would also make the optical and near-IR lines 
too weak. We do not assign an explicit error in the 
estimated $\Ti44$ due to dust cooling. (See also Sect. 4.2.) 

L99 discussed uncertainties in the modeled line fluxes due to the explosion 
model used. As in L99 we have used the 10H explosion model (Woosley \& Weaver 
1986; Woosley 1988), mixed by Kozma \& Fransson (1998ab) to 
give good agreement between their modeling and late time observations. 
Kozma \& Fransson (1998b) compared the results from this model with 
a similarly mixed version of the 11E1 model (Shigeyama et al. 1988), 
and found that the iron lines are insensitive to the explosion model used. 
This is because the iron core mass, which is fixed by the amount of 
ejected $^{56}$Ni, is the same in both models. We follow L99 and assign 
a $15\%$ error in $\Ti44$ due to the choice of explosion model.

In our calculations we have assumed a local deposition of positron energy
from the radioactive decay of $^{44}$Ti. This assumption was also used
in L99. The argument is that the efficiency of trapping cannot change 
substantially until day $\sim 3\,600$ (K00), and that the most straightforward 
interpretation of a near-constant efficiency of trapping over an extended 
period of time is to assume full trapping. L99 assumed a $15 \%$ error 
in $\Ti44$ due to the uncertainty of trapping. Here we do not infer an 
explicit error to the models in Sect. 3, but in Sect. 4.2 we investigate 
this in greater detail in terms of a consistency check of the modeled IR and 
optical fluxes. The positron deposition is discussed in Sect. 4.2
in conjunction with clumping. Even without positron leakage, 
clumping of the iron-rich gas could cause an error in the estimate of $\Ti44$. 
However, according to L99, this error is likely to be small, $\sim 5\%$, and 
can be ignored when compared to other errors discussed above.

The combined error of $f_{26 \mum}$ due to model approximations, 
except for different clumping and positron leakage scenarios (which will 
be discussed in Sect. 4.2)
is therefore $\sim 36\%$. With the line profile used in Fig. 2 we thus 
arrive at an upper limit on $\Ti44$ which is $\sim 8 \EE{-5} \Msun$.

\subsection{Consistency check of the modeled optical and infrared emission}
Although the limit on $\Ti44$ we found above in Sect. 4.1 is a factor 
of $\sim 2$ lower than the limit found by L99 for the $3\,999$ data, 
it is still compatible with K00 (see Sect. 1). However, the 
agreement is not perfect although the same computer code has been used.
Here we make a consistency check of the modeled optical and IR emission
produced by our model. For the late optical photometry (day $3\,268$) 
we adopt the same data as used by K00 (Soderberg, Challis \& Suntzeff 1999). 
The observational error of these is approximately $\pm 0.05$ 
magnitudes, as judged from the systematic error in the zero point of HST/WFPC2 
photometry (Mould et al. 2000; P. Challis 2000, private communcation). 
 
We have concentrated our consistency check on models 
with $\Ti44 = 10^{-4} \Msun$. This is close to the upper limit found
in Sect. 4.1 from the far-IR lines when the errors have been considered. This
value of $\Ti44$ is also in the lower range of the preferred values found 
by K00 in her modeling of the optical and near-IR emission. This can be
seen from Table~2 where the modeled $V$, $R$ and $I$ magnitudes in 
model M1 are all slightly fainter, by roughly half a magnitude, than
the observed, while the predicted [Fe~II]~26~$\mum$ line emission is somewhat 
too strong to be accommodated by our error analysis in Sect. 4.1. 
(Model M1 is identical to the model with $\Ti44 = 10^{-4} \Msun$ and no 
photoionization in Table~1.)

It is obvious that our models contain simplifications that can make us
systematically overproduce far-IR emission at the expense of optical 
emission. We still have to explore fluorescence (see Sect. 3.1), and other
simplifications could be clumping and different scenarios 
for the deposition of positron energy. These two effects are probably also 
coupled. In Sect. 4.1 we mentioned that L99 found that the 
estimated error in $\Ti44$ due to clumping alone should be small. This is 
true if one looks at the total emission coming out from the core, and if 
the energy deposition is local. However, if we assume that the positron 
optical depth scales linearly with density (ignoring effects due to
the magnetic field), positron leakage from the low-density component is
more likely than from the high-density one.
To test this we have, similar to what we did in L99, used a two-component model 
with equal mass in the dense and the less dense components. For the density 
ratio we use the value 5. From a purely geometrical point of view, positrons 
produced in a high-density clump see an optical depth for absorption in the 
positron emitting clump which is $\sim 3$ times higher than for self-absorption 
of positrons emitted in a low-density region. It is therefore more likely that
positrons from the low-density gas leak from their production site, and the 
chance that they are captured in either a low-density, or high-density
component on a larger scale is about equal. From a physical point of view,
absorption of the leaking positrons is slightly biased toward absorption in
the low-density gas since this is more ionized (due to less efficient
recombination), which results in slightly higher deposition there. We have 
tested three scenarios, all without photoionization and 
with $\Ti44 = 10^{-4} \Msun$: the first (M2 and M3) is the same as in L99 
with local deposition of the positron energy. The second (M4 and M5) is with 
all positron deposition (i.e., from both the low and high-density gas) 
occurring only in the high-density clumps. This situation is an extreme
case (cf. above), but could be possible if magnetic effects make deposition
in the high-density gas more likely than in the low-density gas. The third 
(M6 and M7) scenario is for a situation where the positrons in the low-density
gas are instead assumed to leak into and deposit their energy in the the 
silicon-rich gas which is macroscopically mixed with the iron-rich gas. 
We have run these models both as steady state and time dependently to separate 
this effect as well. We have summarized the results in 
Table~2, where we show $f_{24 \mum}$, $f_{26 \mum}$, $V$, $R$ and $I$ 
at 3\,425 days. (The observed optical fluxes in Table~2 are for 
3\,268 days, and not for 3\,425 days. The change in modeled optical flux 
between these two epochs is, however, negligible.)

\begin{table}
\caption[]{Modeled luminosities and magnitudes at 3\,425~days$^{a}$}
\label{modelfluxes_2}
\[
\begin{tabular}{lcccccc}
\hline
\noalign{\smallskip}

Model & Time dep.$^{b}$ & $f_{24\mum}$$^{c}$ & $f_{26\mum}$$^{c}$ & $V$ & $R$ & $I$ \\

\noalign{\smallskip}
\hline
\noalign{\smallskip}

M1$^{d}$ &  yes  & 0.32 & 1.18 &  20.4  & 19.7   & 20.0  \\
M2$^{e}$ &  yes  & 0.39 & 1.30 &  20.3  & 19.7   & 20.0  \\
M3$^{e}$ &  no   & 0.35 & 1.27 &  20.3  & 20.4   & 20.2  \\
M4$^{f}$ &  yes  & 0.47 & 1.13 &  20.1  & 19.6   & 19.9  \\
M5$^{f}$ &  no   & 0.46 & 1.15 &  20.1  & 20.4   & 20.2  \\
M6$^{g}$ &  yes  & 0.32 & 0.80 &  19.6  & 19.5   & 19.7  \\
M7$^{g}$ &  no   & 0.30 & 0.82 &  19.6  & 20.0   & 19.9  \\

\noalign{\smallskip}
\hline
\end{tabular}
\]

$^{a}$$\Ti44 = 10^{-4} \Msun$. No photoionization. Observed 
magnitudes (at 3\,268 days) are $V = 20.0$, $R = 18.9$, and $I = 
19.3$ (Soderberg et al. 1999, P. Challis 2000 [private communcation]).\\
$^{b}$Time dependent or steady state model.\\
$^{c}$Peak flux in Jy for the line profile used in Figs. 1 and 2. Observed
upper limits on $f_{24\mum}$ and $f_{24\mum}$ are 0.39 Jy and 0.64 Jy, 
respectively (see Sect. 1).\\ 
$^{d}$One-component model, i.e., same as in Table 1.\\
$^{e}$Two-component model. Positron deposition in both density components.\\
$^{f}$Two-component model. Positron deposition only in the high-density component.\\
$^{g}$Two-component model. Positron deposition in the high-density component 
and in the Si-rich gas (see text).\\

\end{table}

From the results in Table~2 it is clear that the far-IR emission is 
rather robust to density inhomogeneities in the core, as well as to the 
assumption of steady state as long as the positrons are all trapped
in the iron-rich gas. Allowing for positron diffusion into the Si-rich 
gas decreases the [Fe~II]~25.99~$\mum$ emission; instead the emission in $V$
increases notably, while $R$ and $I$ are less affected. In our most
extreme model M6 (cf. Table~2) the positron escape leads to a reduction 
of $f_{26 \mum}$ by $\sim 32\%$. This model actually produces 26~$\mum$ 
emission which is below the detection threshold, so this shows that for 
extreme situations of clumping and positron depositions our model could
accommodate up to $\Ti44 \sim 1.1\EE{-4} \Msun$ in SN 1987A, and still not 
produce detectable far-IR emission. However, such a scenario would 
imply $V \sim 19.5$ at 3\,425~days, a factor $\sim 1.6$ higher
flux in $V$ than observed. 
As the optical flux is only a minor contribution to the total emission 
coming out from the supernova at this epoch, and we yet have to investigate
fluorescence to model the optical emission in detail, we cannot exclude
a scenario like this. We therefore base our upper limit on the 26~$\mum$ 
line which we believe we have modeled more accurately. To estimate a 
conservative upper limit on $\Ti44$, taking into account various clumping 
and deposition effects, we thus arrive at an upper limit of $1.1\EE{-4} \Msun$. 

\subsection{Considerations of the derived mass of $^{44}$Ti}
Models for the yield of $^{44}$Ti give quite different results. In particular, 
in the model of Timmes et al. (1996; see also Woosley \& Weaver 1995) with 
a zero-age mass of $\MZA = 20 \Msun$ (i.e., corresponding to SN 1987A) the 
mass of the initially ejected $^{44}$Ti is $1.2\EE{-4} \Msun$, 
but only $1.4\EE{-5} \Msun$ escape after fallback. This is in accord with
our upper limit in Sect. 4.1, but it should be emphasized that the variation 
of $\Ti44$ with $\MZA$ in Timmes et al. (1996) is complex, and for models 
with $\MZA = 18 \Msun$ and $22 \Msun$, the calculated $\Ti44$ 
is closer to our limit. Furthermore, the recent models of Hoffman et al. 
(1999) for 15 and 25 solar mass models, indicate that the yield of $^{44}$Ti 
in Woosley \& Weaver (1995) may have to be increased. 

The models of Thielemann et al. (1996) have larger entropy and thus more
alpha-rich freeze-out than those of Woosley \& Weaver (1995). Accordingly, the
ratio $\Ti44 / \Nii$ (where $\Nii$ is the mass of ejected $^{56}$Ni that
does not fall back) is higher. For example, in their $20 \Msun$ 
model $\Nii \approx 0.074 \Msun$ and $\Ti44 \approx 1.7\EE{-4} \Msun$.
This value for $\Ti44$ is higher than our upper limit in 
Sect. 4.2. With the refinements made by Hoffman et al. (1999) to the models
of Woosley \& Weaver (1995), it seems that our upper limit is lower than, or 
at least close to, $\Ti44$ in the models of both groups. The calculated 
$\Ti44$ is, however, very model dependent, and sensitive to, e.g., the 
explosion energy, which in the case of SN 1987A is still only known to an 
accuracy of $\sim 30\%$ (Blinnikov et al. 2000).

The models of Nagataki et al. (1997, 1998) are related to those of Thielemann 
et al. (1996), although they also allow for 2-D. With no asymmetry, 
Nagataki et al. obtain $\Ti44 \sim 6\EE{-5} \Msun$
for a SN 1987A-like explosion, which is a factor of $\sim 3$ lower than
Thielemann et al. (1996), and could indicate the range of uncertainty of the 
modeling. In the models of Nagataki et al. the yield of $^{44}$Ti quickly 
increases with the degree of asymmetry. For an asymmetry of 2 between the 
equator and the poles, the stronger alpha-rich freeze-out in the polar
direction increases $\Ti44$ to $\Ti44 \sim (1.2-1.9) \EE{-4} \Msun$ 
(for $\Nii \approx 0.07 \Msun$), the range of titanium yield depending on the 
form of the mass cut (Nagataki 2000). Nagataki (2000) argues that this degree 
of asymmetry gives a distribution of ejected $^{56}$Ni which agrees with 
observed line profiles of [Fe~II]~1.26~$\mum$ (Spyromilio et al. 1990) and 
[Fe~II]~18~$\mum$ (Haas et al. 1990) at $\sim 400$ days. 

Although the models of Nagataki (2000) thus hint that asymmetry can explain the
distribution of the ejected $^{56}$Ni in SN 1987A, the need for asymmetry
to reach $^{44}$Ti in excess of $10^{-4} \Msun$ cannot be considered a strong
case, given the uncertainty in the modeling. We therefore do not 
regard our upper limit on $\Ti44 \sim 1.1\EE{-4} \Msun$ for SN 1987A 
as contradictory when compared to the absolute yield of $^{44}$Ti in, 
e.g., Nagataki (2000), especially since the models of Nagataki are trimmed 
to produce $\Ti44$ in agreement with the light curve results of Mochizuki \& 
Kumagai (1998).
 
Table 2 also shows that although steady-state models can be used for the
far-IR emission from the core, they fail to reproduce the optical 
broad-band emission. In particular, they underproduce $R$ and $I$ 
magnitudes. For the cases in Table 2, the maximum errors are 0.8 and 0.3 
magnitudes, respectively. This is due to the freeze-out effect 
described by Fransson \& Kozma (1993). Consequently, any estimate 
of $\Ti44$ based on steady-state calculations for the optical broad bands
is likely to overestimate $\Ti44$, even though the structure and 
atomic data are correct. It could be that the models of Mochizuki \& 
Kumagai (1998; see also Nagataki 2000) suffer from this, which may
explain why their lower limit, $\Ti44 = 1.0\EE{-4} \Msun$, almost coincides
with our upper limit, and why their lower limit is higher than the one 
of K00. 

Finally, we note that our upper limit on $\Ti44$ could be higher if dust
contributes significantly to the cooling at late times. This
is not unique to our models, but affects all models trying to reproduce
the optical and IR emission. The range of $\Ti44$ bracketed by K00 
and our analysis, $\Ti44 = (0.5-1.1)\EE{-4} \Msun$ would then be shifted 
to a range with higher masses.
Heavy dust formation in the core 
would, however, block out the line emission from the core and affect line 
profiles. To test the importance of dust cooling, direct measurements 
of the gamma-ray emission at 1.157 MeV, the result of the radioactive 
decay of $^{44}$Ti, are needed. Future, and even more sensitive gamma-ray 
instruments than INTEGRAL, are needed for such a study.

\section{Conclusions}
We have extracted and modeled ISO/SWS data for day 3\,425 of SN 1987A. 
The supernova was neither detected in [Fe~I]~24.05~$\mum$ nor in 
[Fe~II]~25.99~$\mum$, which are the lines expected to be the strongest 
in the spectrum. For a likely line profile, we have used this finding 
to put an upper limit on the mass of ejected $^{44}$Ti, $\Ti44$. 
Including possible modeling errors, we obtain $\Ti44 \lsim 1.1\EE{-4} \Msun$.
This supersedes the result of Lundqvist et al. (1999), who found a 
somewhat higher limit analyzing other data. The new limit is, however, more 
reliable in that we have included more possible sources of error, 
as well as late optical broad band data. It is shown that time dependent 
effects are important for the latter, probably explaining why our 
models give a somewhat lower preferred value of $\Ti44$ compared to 
those in previous steady-state analyses.

\begin{acknowledgements}

We thank Peter Challis, Nikolai Chugai and Stan Woosley for discussions. 
We are grateful to support from the Swedish Natural Science Research Council
and the Swedish National Space Board.

\end{acknowledgements}

{}

\end{document}